\documentclass[letter]{aa} 

\usepackage{graphicx}
\usepackage{txfonts}

\newcommand{\hi}{\mbox{H{\,\sc i}}}

\newcommand{\kms}{km s$^{-1}$}

\newcommand{\msol}{\rm M$_\odot$}

\begin{document}
 
 \title{Galaxy interactions in loose galaxy groups:\\KAT-7 and VLA \hi\ Observations of the IC 1459 group \\
  }
 
 \titlerunning{KAT-7 and VLA \hi\ Observations of IC 1459}
 \authorrunning{Oosterloo et al.} 
 
 \author{T.A. Oosterloo\inst{\ref{inst1}}\fnmsep \inst{\ref{inst2}}\thanks{oosterloo@astron.nl}\and M.-L. Zhang\inst{\ref{inst1}}\fnmsep \inst{\ref{inst3}}\and D. M. Lucero\inst{\ref{inst2}}\fnmsep \inst{\ref{inst4}}\fnmsep \inst{\ref{inst5}}\and C. Carignan\inst{\ref{inst4}}\fnmsep \inst{\ref{inst6}}}
          
  \institute{
  ASTRON, Netherlands Institute for Radio Astronomy, Postbus 2, 7990 AA Dwingeloo, The Netherlands\label{inst1}
  \and
   Kapteyn Astronomical Institute, University of Groningen, Postbus 800, 9700 AV Groningen, The Netherlands\label{inst2}
  \and
  Leiden Observatory, Leiden University, Postbus 9513, 2300 RA Leiden, The Netherlands\label{inst3}
  \and
  Department of Astronomy, University of Cape Town, Private Bag X3, Rondebosch 7701, South Africa\label{inst4}
  \and
  Department of Physics, Virginia Tech, 850 West Campus Drive, Blacksburg, Virginia 24061, United States \label{inst5}
  \and
  Observatoire d'Astrophysique de l'Universit\'e de Ouagadougou, BP  7021, Ouagadougou 03, Burkina Faso\label{inst6}
                } 

\abstract
{We report on the results from deep \hi\ observations, performed with the Karoo Array Telescope  and with the Karl G.\ Jansky Very Large Array of the loose galaxy group centred on the early-type galaxy IC 1459. 
The main result from our observations is the detection of a nearly continuous, 500-kpc long \hi\ tail which crosses the entire group. Earlier observations with the Australia Telescope Compact Array   had shown the presence of  a large \hi\ tail in this galaxy group, but because of the much larger  coverage of the new data, the full extent of this tail is now visible. The \hi\ mass of this structure is $3.1 \pm 0.3 \times 10^9$ \msol.    Based on its morphology and kinematics, we conclude that the  tail consists of gas stripped from NGC 7418 through tidal interactions, with ram-pressure affects playing at most a minor role. Optical images of the IC 1459 group do not show many indications that galaxy interactions are common in this group. The \hi\ data reveal a very different picture and  show that almost all gas-rich galaxies in the IC 1459 group have a distorted \hi\ distribution indicating that many interactions are occurring in this group. This high number of interactions shows  that the processes that drive galaxy transformation are also occurring in fairly loose galaxy groups. }


\maketitle
%

\section{Introduction}
\label{sec:intro}

\hi\ observations offer an effective diagnostic tool for tracing  interactions between galaxies or between galaxies and their large-scale environment. Spiral galaxies near the central regions are often found to be \hi\ deficient \citep{dav73, cha80, gio85}, because of ram-pressure stripping by the hot intracluster medium (ICM)  as well as due to tidal interactions.
Direct evidence for these processes has been found in many galaxy clusters in the form of long gas tails .
Both the   neutral as well as the ionised hydrogen  in galaxy disks can be affected by these interactions  \citep[e.g.,][]{oos04,oos05,che06,chu07,chu09,gul17}.


As shown many years ago by the discovery of the morphology--density relation \citep{dre80, pos84}, it is such interactions that drive the evolution from late to early type and from gas-rich to gas-poor galaxies \citep[e.g.][]{cay90}. However, the morphology--density relation exists also outside  cluster environments \citep{cap11,ser12} and there is a smooth transition of the mix of galaxy types going from the lowest galaxy density  environments, via groups of galaxies, to high galaxy density environments of rich clusters. Therefore, the processes that change galaxy morphology must also act outside clusters, albeit to a lesser degree. 

For nearby galaxy groups,  it is not easy to study the role of galaxy interactions using \hi\ observations, because such groups are often much larger than the field of view of current radio interferometers and large amounts of observing time are needed to make a full inventory of the \hi\ in such groups. However, evidence for interactions being relevant is now appearing through, for example,  the study of the \hi\ properties of galaxies in compact groups \citep[e.g.,][]{ver01},  the detection of large \hi\ tails in galaxy groups, such as the recent detection of a large \hi\ tail in and around the compact group HCG44 \citep{ser13, hes17} or the numerous tails observed in the region of the Leo Cloud \citep{lei16}. 

The large-area \hi\ surveys planned  for upcoming instruments like ASKAP, Apertif, Fast, and MeerKat will greatly increase the number of groups for which one can study in detail the role of interactions.
In this paper, we report  on observations that have been done as part of the preparation for these large upcoming surveys. These observations show  that  the processes  that drive galaxy transformation also occur in much less massive structures, such as galaxy groups, and that \hi\ observations are particularly well suited to trace these processes. Earlier  observations with the Australia Telescope Compact Array (ATCA)  of the galaxy group surrounding the early-type galaxy IC 1459 had shown the presence of \hi\ tails  and other signs of interactions such as distorted \hi\ morphologies \citep{oos99}. This contrasts with what is discerned from optical images, which show only few indications that interactions between group members are occurring. The group is  relatively young with an early-to-late types ratio of only 0.29 \citep{bro06}. The \hi\ observations by \citet{sen07} also show signs of ongoing interactions, while it was also found   that many galaxies in the group have shrunken \hi\ discs compared to  field galaxies.
Recently, \citet{ser15}, using six antennas of the Australian Square Kilometre Array Pathfinder (ASKAP), imaged the \hi\ in 11 galaxies down to column densities $\sim10^{20}$ cm$^{-2}$ inside a $\sim6$ deg$^2$ field. Also these data  show that interactions are common in this group. More recent ATCA observations by \citet{sap17} discussed the debris to the SW and SE of IC 1459 and concluded they account for a large fraction of the IGM. 

\begin{figure}
\includegraphics[width=\columnwidth]{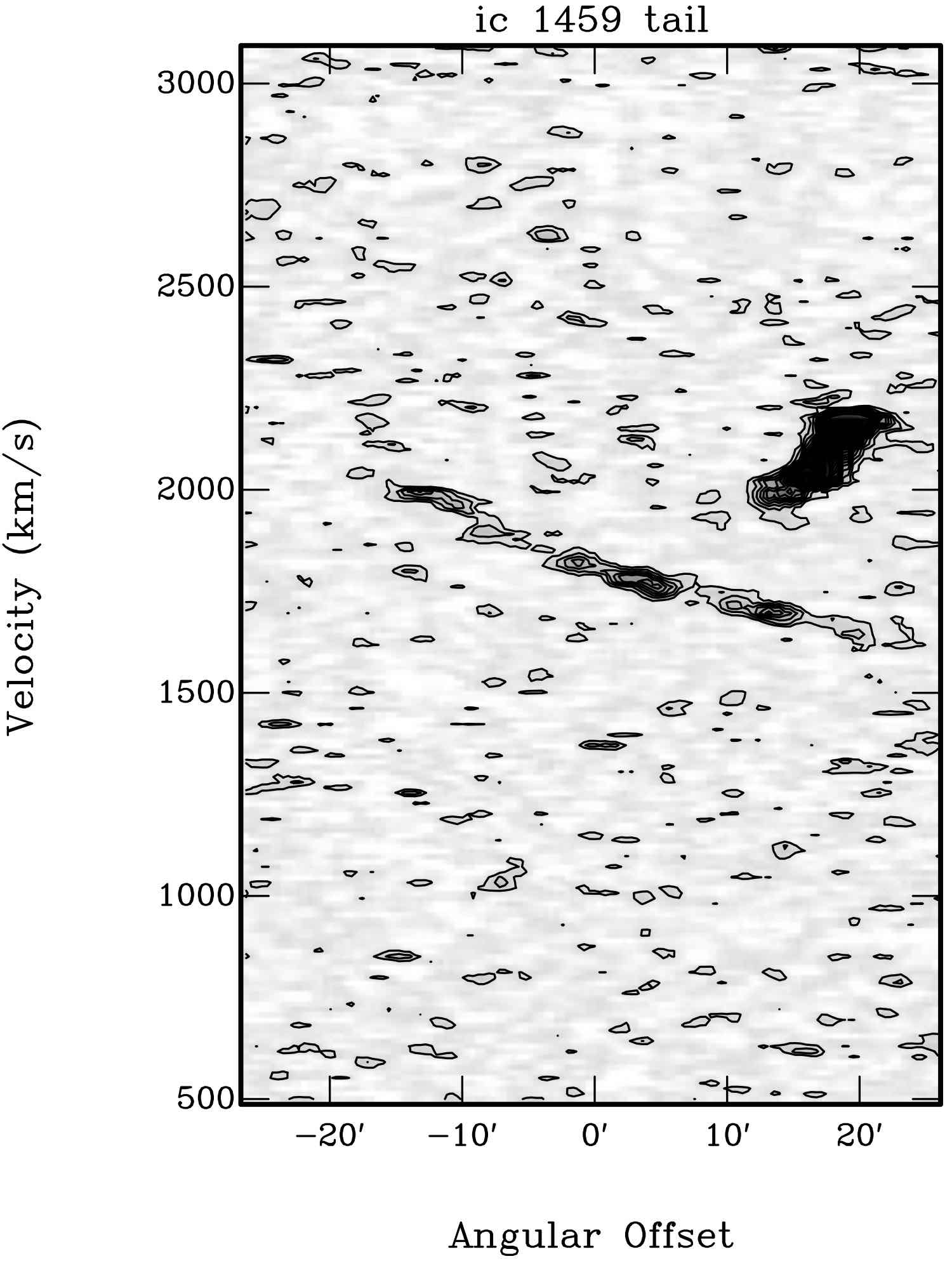}
\caption{Position-velocity slice taken from the ATCA observations by \citet{oos99} along PA = $30\deg$ along the \hi\ tail centred on NGC 7418A, showing   a gap of $\sim250$ \kms\ between the gas of NGC 7418A and the tail.}
\label{fig:oldATCA}
\end{figure}

Most interestingly,  the ATCA data  of \citet{oos99} show a long ($>$200 kpc) low surface brightness \hi\ tail running from NE of IC 1459 to SW of it. However, a puzzling feature is that this  tail appears to be connected to NGC 7418A spatially, but it is not connected to this galaxy in velocity, with a gap of $\sim$250 km s$^{-1}$ (Fig.\ \ref{fig:oldATCA}). Here we report on deeper \hi\ observations,  obtained with the Karoo Array Telescope (KAT-7: the South African SKA pathfinder)  radio telescope  \citep{car13, luc15, fol16}, which cover the entire group to lower column density and which reveal the full extent of this intra-group \hi. We also report on higher-resolution follow-up observations done with the Karl G.\ Jansky Very Large Array (VLA; CnB \& D arrays)  centred on NGC 7418 and NGC 7418A to examine one of the \hi\ tails in more detail.


 The remainder of this paper is as follows. In Sec.~\ref{sec:IGM}, we present the two data sets, followed  by a discussion of the cool IGM in the group. In Sec.~\ref{sec:disc} we try to understand the origin of those \hi\ structures between the group members and discuss the implications of those structures on the evolution of such loose groups. Finally,  Sec.~\ref{sec:concl}, we summarise our results.  A distance of 29 Mpc \citep{ser15} is adopted in this study for this group implying a scale of 1 arcminute corresponding to 8.4 kpc.


\section{Intergalactic \hi\ in the IC 1459 group}
\label{sec:IGM}

The parameters of the KAT-7 and VLA data are given in Table \ref{obspar}. The KAT-7 data consist of a mosaic of 2 fields in order to cover most of the galaxy group. Although KAT-7 consists of only seven 12-m dishes, its compact array configuration and the excellent receivers make it  very sensitive for extended, low-surface brightness emission. The VLA data were used to cover a single high-resolution field around NGC 7418 and NGC 7418A, in order to get higher resolution information of \hi\ around the galaxy NGC 7418. The KAT-7 data have a spatial resolution of $3.5\arcmin$, a velocity resolution of 12.8 \kms and a RMS noise of 1.5 mJy beam$^{-1}$, while the VLA data have a beam of $97\arcsec \times 51\arcsec$, a velocity resolution of 20 \kms and a RMS noise of 0.3 mJy beam$^{-1}$. 

\begin{table}
\centering
\caption{Parameters of the KAT--7 \& VLA observations.}
\label{obspar}
\begin{tabular}{lr}
\hline\hline
KAT-7 & \\
\hline
Total integration  &  60 hours\\
FWHM of primary beam & 1.08$^{\circ}$  \\
Total Bandwidth & 13 MHz \\
Central frequency &  1411.9 MHz \\
Number of channels & 300 \\
Velocity resolution (after Hanning) & 12.8 \kms \\
Weighting function & Robust = 0.5 \\
FWHM of synthesized beam & 213\arcsec  x 207\arcsec \\
RMS noise  & 1.5 mJy/beam \\
Column density limit& \\
(3$\sigma$ over 16 km s$^{-1}$) & $1.8 \times 10^{18}$ cm$^{-2}$ \\
\hline
VLA & \\
\hline
Total integration  (D array) &  15.8 hours\\
Total integration  (CnB array) &  5.65 hours\\
FWHM of primary beam & 32\arcmin \\
Total Bandwidth & 32 MHz \\
Central frequency &  1403.9 MHz \\
Number of channels & 250 \\
Velocity resolution (after Hanning) & 20.0 \kms \\
Weighting function & Robust = 2\\
FWHM of synthesized beam & 97.2\arcsec  x 50.6\arcsec \\
RMS noise  & 0.3 mJy/beam \\
Column density limit& \\
(3$\sigma$ over 16 km s$^{-1}$) & $3.3 \times 10^{18}$ cm$^{-2}$ \\
\hline\hline
\end{tabular}
\end{table}

\begin{figure}
\includegraphics[width=0.95\columnwidth]{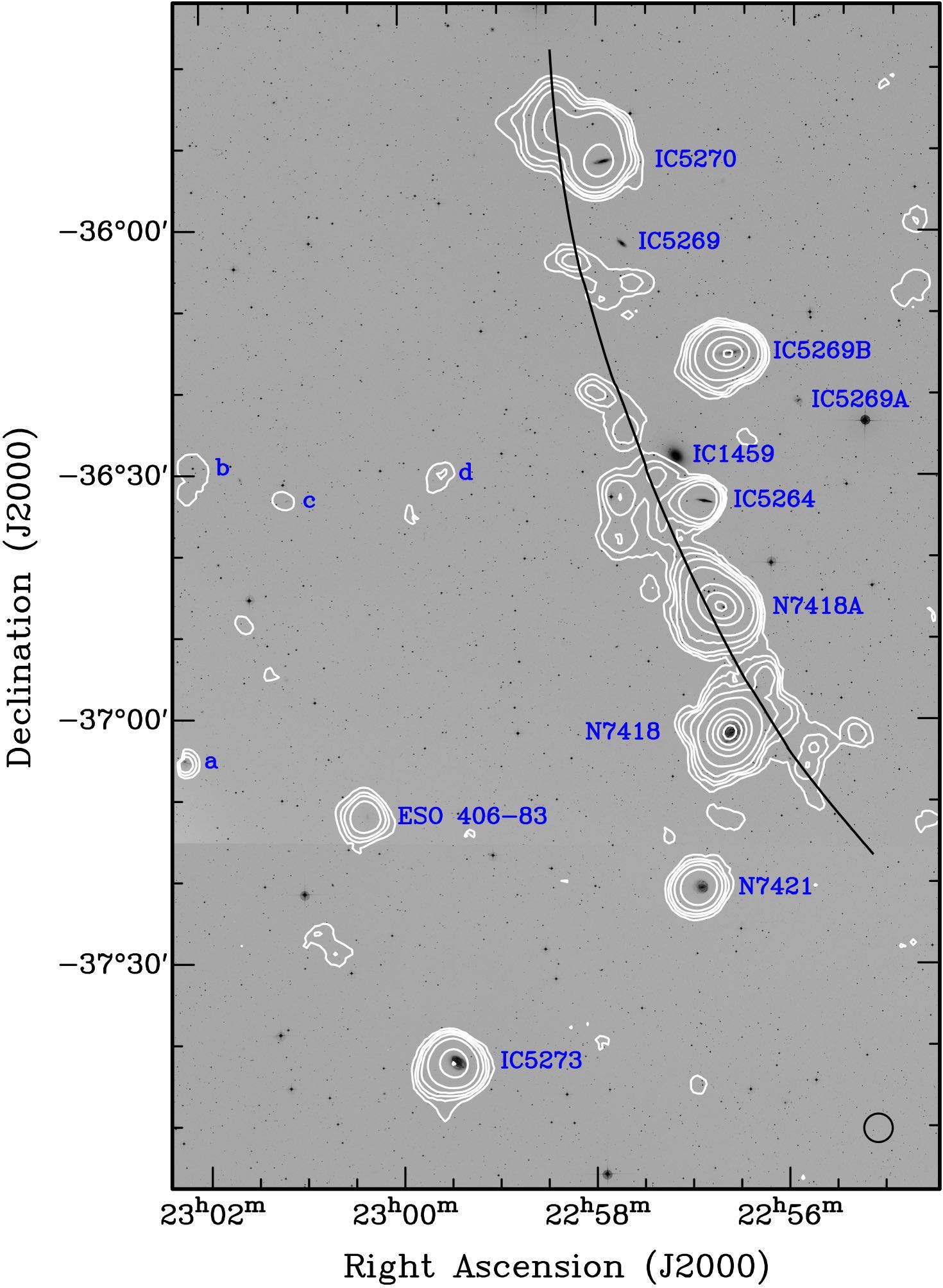}
\caption{Total \hi\ map from the KAT-7 observations. The galaxy denoted by a is  ESO 406-G42, b is  DUKSR 406-83, c is  GALEX ASC J230109.65-363307.3 and d is  2MASX J22593416-3629514. The beam is shown in the bottom-right corner.  Contour levels are  5.0, 10, 20, 50, 100 and 500 $\times 10^{18}$ cm$^{-2}$. Also shown is the locus used for     Fig.\ \ref{fig:tails}. }
\label{fig:KAT7mom0}
\end{figure}

The main result from our observations is the detection of a nearly continuous \hi\ tail which crosses the entire group, starting north of IC 1459  to beyond NGC 7418, south of IC 1459 (Fig.\ \ref{fig:KAT7mom0}).    The  ATCA data of \citet{oos99} had shown the presence of  a large \hi\ tail in the IC 1459 group, but because of the much larger KAT-7 coverage, the full extent of this tail is now visible.  The ATCA data suggested a possible link with NGC 7418A, but the KAT-7 image shows that this is not the case. North of the continuous part of the tail, in the direction of IC 5270, a large \hi\ cloud is observed which may form a single structure together with the continuous tail. The \hi\ tail extends nearly $1\deg$, which corresponds to $\sim$ 500 kpc. From the KAT-7 data, we find that the  \hi\ flux integral  of  this tail is $15.7 \pm 1.6$ Jy \kms, which corresponds  to an \hi\ mass of $3.1 \pm 0.3 \times 10^9$ \msol. The morphology of the tail, and the fact that it appears to connect to NGC  7418, suggests that it consists of \hi\ stripped from this galaxy. However, the tail extends beyond NGC 7814 which makes it unlikely the tail is due to ram-pressure stripping. 

\begin{figure}
\includegraphics[width=\columnwidth]{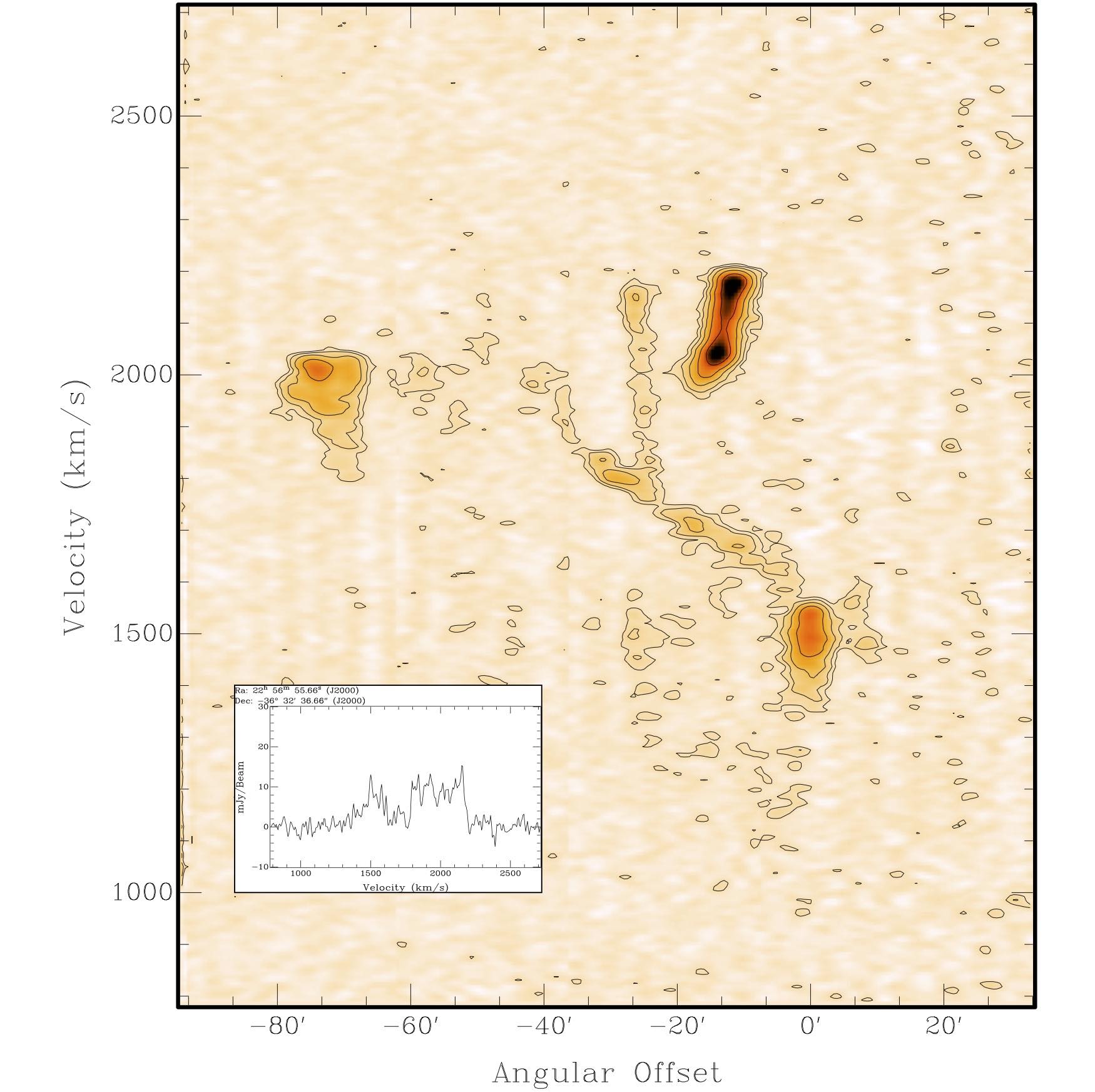}
\caption{PV diagram taken along the path shown in Fig.\ \ref{fig:KAT7mom0} showing the velocity structure of the long \hi\ tail. The inset is the spectrum taken at offset -27$^\prime$ corresponding to IC 5270 and which shows the very wide profile detected at that position. }
\label{fig:tails}
\end{figure}

Figure \ref{fig:tails} shows a position-velocity diagram of the large \hi\ tail taken along the curved path shown in Fig.\ \ref{fig:KAT7mom0}. This figure shows that also the kinematics of the tail suggest a link with NGC 7418. However, Fig. \ref{fig:tails} also shows that the northern part of the tail, the cloud N of the tail, and IC 5270 lie at similar velocities, perhaps indicating that IC 5270 also may have a connection to the intragroup \hi\ observed.  The very asymmetric \hi\ distribution of IC 5270 reinforces this \citep[see also][]{ser15}.
 
In addition to the large intragroup \hi\ structure,  there is ample evidence of several other interactions taking place and the IC 1459 group is much more lively than one would expect based on optical images. From Fig.\  \ref{fig:KAT7mom0}, one can see that most  galaxies detected have an \hi\ distribution which is (very)  asymmetric. Starting from the north, the \hi\ disk of IC 5270 is quite extended to the NE (as was also observed by \citealt{ser15}), IC 5269B to the SE, IC 5264 to the E, NGC 7418A (the only galaxy showing optical distortions) to the NE, NGC 7418 to the SW, ESO 406-83 to the S, NGC 7421 to the SE and IC 5274 to the SE. In fact, none of the gas-rich members have a regular well centred \hi\ distribution. The same is true for the 4 small galaxies to the east of the field:  ESO 406-G42, DUKSR 406-83,  Galex ASC J230109.65-363307.3, and  2MASX J22593416-3629514. 

Another feature, not evident from the integrated \hi\ distribution, is revealed in Fig.\ \ref{fig:tails}, when looking at the spectrum taken at a position near IC 5264. \hi\ is detected over a very large velocity range of 800 \kms\ and the profile seems to consist of the two-horned profile from IC 5264 with in addition an \hi\ cloud at lower velocities.
No optical counterpart is present to this cloud and one possibility is that IC 5264 is moving mostly along the line of sight and that the \hi\ cloud is a tail originating from IC 5264 and which is seen edge on.

\begin{figure}
\includegraphics[width=\columnwidth]{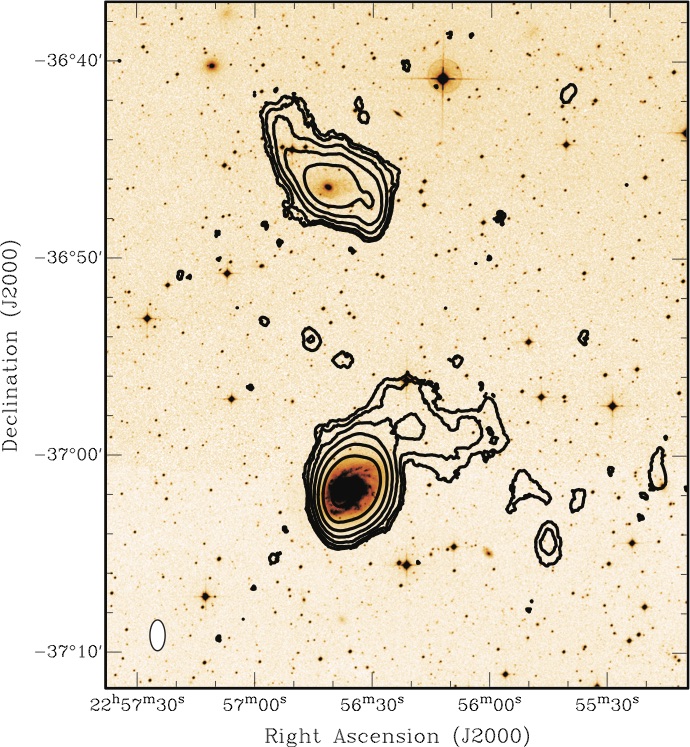}
\caption{VLA data: total \hi\ map from the VLA observations of NGC 7814 (bottom) and NGC 7418A (top). The beam is shown in the bottom-left corner. Contours are 1.2, 3.0, 6.0 12.0 $ \times 10^{19}$ cm$^{-2}$.}
\label{fig:JVLAmom0}
\end{figure}

The structure of the large \hi\ tail near NGC 7418 is complex and to further investigate the origin of the large \hi\ tail, and its connection to NGC 7418, we performed higher-resolution observations with the VLA centred on NGC 7418. Figure \ref{fig:JVLAmom0} shows the integrated \hi\ emission resulting from this observation. In the VLA data, we only detect that part of the \hi\ tail which is west of NGC 7418 and the morphology clearly shows that it consists of \hi\ coming from NGC 7418. The larger, eastern part of the tail is not (or barely) detected in the VLA observations. This is consistent with the KAT-7 data, which show that towards NGC 7418, the eastern tail gets fainter and that it may consists of more diffuse \hi. For the tail pointing W from NGC 7418 we measure a flux integral of $3.9 \pm 1.5$ Jy \kms\ from the VLA data and of $6.0 \pm 1.5$ Jy \kms from the KAT-7 data, which corresponds to an \hi\ mass of $1.2 \pm 0.3 \times 10^9$ \msol.

\section{Discussion }
\label{sec:disc}

One of the main questions coming from the data presented is    the possible origin of the large \hi\ features observed. A number of mechanisms can create large \hi\ tails, such as tidal interactions between galaxies, or  between a galaxy and a galaxy group as a whole, and ram-pressure stripping due to an intracluster or intragroup gaseous medium.   If one assumes that the large \hi\ tail east of NGC 7418 and the western tail are both   related to this galaxy,  and that they have a common origin, one could conclude that  they have a tidal origin because such twin-arms are characteristic for tidal interactions. The most likely  candidate that NGC 7418 is interacting with is NGC 7418A, a galaxy which also shows strong signs of tidal interactions, even on  optical images.
This morphology of the \hi\ structure is very different from what is observed for one-sided \hi\ tails in, for example, the Virgo cluster and which are caused by ram-pressure stripping.
 To some extent, the system seen in IC 1459 is, apart from differences in scale,  reminiscent of the structure of the Magellanic Stream system with a long trailing tail (main Magellanic Stream: $\sim$$200$ kpc) and a shorter Leading Arm  ($\sim$$25$ kpc; see \citealt{put98}). The  Magellanic Stream and the Leading Arm are thought to be mainly due to tidal effects, although ram-pressure likely plays a role at shaping the overall structure at some level. The similarity of the Magellanic Stream and Leading Arm with the \hi\ tail in the IC 1459 group may suggest that the origin of the large intragroup \hi\ structures is similar. 
 This may be consistent with results from X-ray observations, which have detected X-ray emission centred on IC 1459 \citep{osm04}. But examination of the XMM and Chandra data shows that the observed extent is of the order of the optical size of IC 1459 and a  hot intragroup medium is not detected. Ram pressure depends on density and velocity as $\rho v^2$. Given the lower density of the hot intragroup medium, and the lower velocities (the velocity dispersion of the IC 1459 group  is $223 \pm 62$ \kms; \citealt{bro06}), ram-pressure is likely to be at least two orders of magnitudes smaller than in the Virgo cluster and is unlikely to play a major role in the IC 1459 group. Given the length of the tail ($\sim$500 kpc) and the velocity dispersion of group, the age of the \hi\ tail must be at least  of order $d/\sigma\simeq 2$ Gyr. This is at least an order of magnitude larger than the estimated ages of the \hi\ tails seen in the Virgo cluster \citep{oos05}, underlining a reduced role for the intragroup medium in evaporating cold intragroup gas which may explain the large age of the tail.

\section{Summary and Conclusions}
\label{sec:concl}

Our observations  clearly show galaxy interactions are playing an important role in the evolution of the galaxies of the IC 1459 group. This would not be evident if deep \hi\ observations were not available for this group. The clearest manifestation is a 500-kpc long \hi\ filament emanating from NGC 7418. In addition, all gas-rich galaxies have distorted \hi\ morphologies to some degree, indicating that galaxy interactions are very common. Tidal interactions are likely the main mechanism for the origin of the distorted \hi\ properties, with ram-pressure stripping playing at most a minor role. This high number of interactions in the fairly loose IC 1459 galaxy group shows  that the processes that drive galaxy transformation are also occurring in these kinds of environments.
In the near future, a number of large-area  \hi\ surveys will be undertaken (Apertif, ASKAP, MeerKat, Fast) which will overcome the current observational limitations to find and to image large, low column density \hi\ structures for a large number of  galaxy groups which will help to improve our understanding of the evolution of galaxies in groups.  


\begin{acknowledgements}

We thank all the teams of SKA South Africa for allowing us to get scientific data during the 
commissioning phase of KAT-7 and the VLA TAC for the allocations of C \& D array observing time. 
CC's work is based upon research supported by the South African Research Chairs Initiative (SARChI) of the Department of Science 
and Technology (DST),  the Square Kilometer Array South Africa (SKA SA) and the National Research Foundation (NRF).
The research of DL has been supported by SARChI, SKA SA, and ASTRON Fellowships. DL acknowledges support by the Vici grant  016.130.338 from the Innovational Research Incentives Scheme of the Netherlands Organisation for Scientific Research (NWO).
\end{acknowledgements}


\end{document}